\documentclass[12pt]{article}
\usepackage{amsfonts}
\usepackage{amssymb}
\usepackage{amsmath}
\usepackage{amsthm}
\usepackage{amsfonts}
\usepackage{multicol}
\usepackage{amscd}
\usepackage{textcomp}
\usepackage{multirow}
\usepackage[english, activeacute]{babel}

\def\be{\begin{equation}}
\def\ee{\end{equation}}
\def\bea{\begin{eqnarray}}
\def\eea{\end{eqnarray}}

\newcommand{\sect}[1]{\setcounter{equation}{0}\section{#1}}

\def\1{\'{\i}}

\def\>#1{{\mathbf#1}}

\def\la{\lambda}

\def\eme{N}

\def\kk{K}

\begin{document}

\thispagestyle{empty}

\hfill \today

\ 
\vspace{0.5cm}

\begin{center}

{\Large{\sc{Integrable perturbations of the\\[4pt]
$N$-dimensional isotropic oscillator}} }

\end{center}

\medskip

\begin{center} \'Angel Ballesteros and Alfonso Blasco 
\end{center}

\begin{center} {\it {Departamento de F\1sica,  Universidad de Burgos, 
09001 Burgos, Spain}}

e-mail: angelb@ubu.es, ablasco@ubu.es
\end{center}

  \medskip

\begin{abstract} 
\noindent
Two new families of completely integrable perturbations of the $N$-dimensional isotropic harmonic oscillator Hamiltonian are presented. Such perturbations depend on arbitrary functions and $N$ free parameters and their integrals of motion are explicitly constructed by making use of an underlying $h_6$-coalgebra symmetry. Several known integrable Hamiltonians in low dimensions are obtained as particular specializations of the general results here presented. An alternative route for the integrability of all these systems is provided by a suitable canonical transformation which, in turn, opens the possibility of adding  $(N-1)$ `Rosochatius' terms that preserve the complete integrability of all these models.

 \end{abstract}

\bigskip\bigskip\bigskip\bigskip



\vfill
\newpage


\noindent

\sect{Introduction}

The harmonic oscillator is indeed the paradigm of integrable Hamiltonian dynamics on the $N$-dimensional (ND) Euclidean space, and its superintegrability properties are very well established through group-theoretic methods (see, for instance,~\cite{Perelomov,Fradkin}). Therefore, the construction of non-trivial integrable perturbations/deformations of the ND harmonic oscillator (understood in a broad sense as the addition of any ND potential term -not necessarily small- that preserves the complete integrability of the system) are interesting from both physical and mathematical viewpoints.

In this respect, although certain 2D and 3D perturbations have been thoroughly studied (see, for instance~\cite{Gutzwiller}-\cite{MRWreports} and references therein) the number of ND results is more limited. Among them we can quote the search for coupled polynomial perturbations~\cite{RamaniNuevos}-\cite{BruschiC}, Garnier systems~\cite{Garnier}, the Smorodinski-Winternitz Hamiltonian~\cite{fris,evans2}, Calogero-Moser systems~\cite{Ca71, Mo75} and coupled oscillators coming from symmetric spaces~\cite{FWM}. From a more generic perspective, integrable oscillators on ND curved spaces can also be considered as deformations of the ND Euclidean ones in which the kinetic energy term is also perturbed through functions of the canonical coordinates containing a curvature parameter (see~\cite{BH07,PhysD,Annals} and references therein).

In this paper we demonstrate the complete integrability of the following ND natural Hamiltonian system
\be
\mathcal{H}_{r}=
 \dfrac{1}{2}\sum\limits_{i=1}^{N}\,p_{i}^{2}+\delta_{1}\,\sum\limits_{i=1}^{N}q_{i}^{2}
+ \mathcal{V}_{-2}\left(\sum_{i=1}^N \la_i  q_i,  \sum_{i=1}^N  q_i^2 \right)
+\mathcal{J}\left(\sum\limits_{i=1}^{N}q_{i}^{2}\right)
\label{Hamr}
\ee
where $\mathcal{V}_{-2}$ is any homogeneous function of degree $-2$ in the canonical coordinates and such that $\mathcal{V}_{-2}$ can be written in terms of the variables $\sum_{i=1}^N \la_i  q_i$ and $\sum_{i=1}^N  q_i^2$. Here,  $\delta_{1}$ and $ \la_i \  (i=1,\dots,N)$ are $(N+1)$ free parameters and $\mathcal{J}$ is an arbitrary radial perturbation.  Since the expressions for the integrals of the motion are analytic in $\delta_1$, the oscillator term can also be removed without altering the integrability properties of the system. As a consequence,  by considering appropriate specializations of the dimension $N$, the arbitrary functions $\mathcal{V}_{-2}$ and $\mathcal{J}$ and the $(N+1)$ free parameters $ \la_i \  (i=1,\dots,N)$, we can obtain many different integrable systems.

In Section 2 we shall prove this result by giving explicitly a set of $(N-1)$ functionally independent integrals of the motion for $\mathcal{H}_{r}$, hat will be  obtained by making use of the underlying $h_6$-Poisson coalgebra symmetry~\cite{BR,BHnonlin,alfonsoh6} of the Hamiltonian $\mathcal{H}_{r}$. In fact, such symmetry implies that this Hamiltonian has two different sets of $(N-1)$ independent integrals~\cite{CRMAngel}, and among both sets we find a total number of $(2N-4)$ independent constants of the motion in involution with $\mathcal{H}_r$, which implies that (\ref{Hamr}) is a superintegrable system (albeit not maximally superintegrable). 

Moreover, another family of completely integrable perturbations of the ND oscillator is introduced in Section 3, namely
\be
\mathcal{H}=\dfrac{1}{2}\sum\limits_{i=1}^{N}\,p_{i}^{2}+\delta_{1}\sum\limits_{i=1}^{N}q_{i}^{2}+ \mathcal{F}\left(\sum\limits_{1\leq i<j}^{N}\left(\lambda_{j}q_{i}-\lambda_{i}q_{j}\right)^{2}\right)
+\mathcal{G}\left(\sum\limits_{i=1}^{N}\,\lambda_{i}q_{i}\right) 
\label{Ham1}
\ee
where $\mathcal{F}$ and $\mathcal{G}$ are arbitrary functions and $\delta_{1}$ and $ \lambda_i \  (i=1,\dots,N)$ are again $(N+1)$ free parameters. Again, the second term $(\mathcal{F}+\mathcal{G})$ can be interpreted as two independent families of integrable deformations/perturbations whose superposition preserves the complete integrability of the whole system. 
The (super)integrability of this family of ND Hamiltonian systems can also be proven by restorting to its $h_6$-Poisson coalgebra symmetry and we shall present a number of known integrable polynomial perturbations that can be easily obtained as particular cases of $\mathcal{H}$.

We would like to mention that in the 3D case Hamiltonians (\ref{Hamr}) and (\ref{Ham1}) are subclasses of the systems considered in~\cite{Rauchh6,woja} through a 3D realization of  a so-called `extended' $sl(2,\mathbb R)$ Poisson-Lie algebra (which is just isomorphic to $h_6$). However, the ND integrable generalization here presented is only possible by endowing $h_6$ with the appropriate underlying coalgebra structure and by imposing the full Hamiltonian to be $h_6$-coalgebra invariant. 

In Section 4 we discuss how the complete integrability of (\ref{Ham1}) can be alternatively explained through the introduction of a ND canonical transformation $(q,p)\rightarrow (Q,P)$ under which the complete integrability of (\ref{Ham1}) comes immediately from both  its separability in terms of the (new) $Q_1$ coordinate and the radial symmetry of the remaining $(N-1)$-dimensional Hamiltonian. Moreover, we show that this radial symmetry can be broken without destroying the complete integrability of the system through the addition of  a `Rosochatius' contribution~\cite{Rosochatius1,Rosochatius2} of the form $\sum\limits_{i=2}^{N}\,{b_i}/{Q_{i}^{2}}$. This fact will be algebraically rephrased through the introduction of a new ND symplectic realization for the $h_6$-Poisson coalgebra that includes the previous Rosochatius terms. As a consequence, a further integrable generalization of the Hamiltonians (\ref{Hamr}) and (\ref{Ham1}) is presented.


\noindent

\sect{A new ND integrable Hamiltonian}

In order to prove the complete integrability of the ND Hamiltonian (\ref{Hamr}) we consider the functions \bea
&& A_+ =\sum_{i=1}^N \la_i p_i \qquad
 A_- =
\sum_{i=1}^N \la_i  q_i\qquad   \kk = \sum_{i=1}^N
\bigg(q_i p_i -\frac {\la_i^2}2\bigg)  \cr 
&& B_+ =\sum_{i=1}^N 
{p_i^2} \qquad\!\quad
 B_- = 
\sum_{i=1}^N  q_i^2\qquad\,\,
 M =\sum_{i=1}^N
\la_i^2   ,
\label{sympn}
\eea 
where $\la_i$ are arbitrary parameters and $N$ fixes the number of degrees of freedom. These six functions provide a ND symplectic realization of the six-dimensional $h_6$-Poisson coalgebra with Lie-Poisson brackets given by~\cite{BHnonlin,alfonsoh6,Gilmore}
\be
\begin{array}{lll}
\{\kk,A_+\}=A_+&\qquad \{\kk,A_-\}=-A_- &\qquad \{A_-,A_+\}=M\cr
\{\kk,B_+\}=2B_+&\qquad \{\kk,B_-\}=-2 B_- &\qquad \{B_-,B_+\}=4\kk+2M\cr
\{A_+,B_-\}=-2 A_-  & \qquad \{A_+,B_+\}=0& \qquad
\{M,\,\cdot\,\}=0\cr
\{A_-,B_+\}=2A_+&\qquad \{A_-,B_-\}= 0.& 
\end{array}
\label{poisson}
\ee
The Casimir functions for $h_6$~\cite{Patera} are given by the central generator $M$ and by
\be
{\cal C}= (MB_+ -A_+^2)(MB_- - A_-^2)-
(M \kk -   A_- A_+ +M^2/2)^2 .
\label{cash6}
\ee

Now, let
\be
\mathcal{V}_{-2}\left(A_{-},  B_{-} \right)=\mathcal{V}_{-2}\left(\sum_{i=1}^N \la_i  q_i,  \sum_{i=1}^N  q_i^2 \right)
\label{menos2}
\ee
be a homogeneous function of degree $-2$ in the canonical coordinates that can be written in terms of the $h_6$ generators $A_{-}$ and $B_{-}$. Then the ND Hamiltonian 
\be
\mathcal{H}_{r}= \dfrac{1}{2}\sum\limits_{i=1}^{N}\,p_{i}^{2}+\delta_{1}\,\sum\limits_{i=1}^{N}q_{i}^{2}
+ \mathcal{V}_{-2}\left(\sum_{i=1}^N \la_i  q_i,  \sum_{i=1}^N  q_i^2 \right)
+\mathcal{J}\left(\sum\limits_{i=1}^{N}q_{i}^{2}\right)
\label{Hr}
\ee
where $\mathcal{J}$ is an arbitrary radial function and $\delta_{1},\lambda_i\  (i=1,\dots,N)$ are free parameters,  is completely integrable.

In order to prove this result, we realize that $\mathcal{H}_{r}$ is $h_6$-coalgebra invariant, since it can be written as a smooth function of the $h_6$-coalgebra generators in the form
\be
\mathcal{H}_{r}=
\dfrac{1}{2}B_{+}+\delta_{1}\,B_{-}+\mathcal{V}_{-2}\left(A_{-},  B_{-} \right) +\mathcal{J}\left(B_{-}\right).
\label{Hrabs}
\ee
This invariance immediately implies (see~\cite{alfonsoh6})  that the following $(N-2)$ functionally independent functions
\be
\mathcal{C}_{h_{6}}^{(m)}=\sum\limits_{i<j<k}^{m}\!\!\left(\!
\lambda_{i}(p_{j}q_{k}-p_{k}q_{j})\!+\! \lambda_{j}(p_{k}q_{i}-p_{i}q_{k})\!+\!\lambda_{k}(p_{i}q_{j}-p_{j}q_{i})
\right)^{2}\, 
\qquad m=3,4,\dots, N
\label{integrals}
\ee
Poisson-commute with $\mathcal{H}_r$ and among themselves. In fact, a given integral $\mathcal{C}_{h_{6}}^{(m)}$ is just the  Casimir function (\ref{cash6}) divided by $M$ and expressed in terms of the corresponding $m$-particle realization of the $h_{6}$ algebra given in the form (\ref{sympn}). 

Therefore, we have to find only one more independent integral in order to get complete integrability. In this case such an additional integral can be found through direct computation by making use of the Poisson brackets (\ref{poisson}), and reads
\be
\mathcal{I}_{r}= B_-B_+-(\kk+\frac 12
M)^2
+2\, B_-\,\mathcal{V}_{-2}\left(A_{-},  B_{-} \right)
\ee
that written in terms of the canonical coordinates gives:
$$
\mathcal{I}_{r}= \sum_{{1\le i<j}}^\eme (q_j p_i -
q_i p_j)^2
+2\, \left( \sum_{i=1}^N  q_i^2 \right)\,\mathcal{V}_{-2}\left(\sum_{i=1}^N \la_i  q_i,  \sum_{i=1}^N  q_i^2 \right).
$$
Again, the involutivity of $\mathcal{H}_{r}$ and $\mathcal{I}_{r}$ with respect to the integrals (\ref{integrals}) is a direct consequence of the $h_6$-coalgebra symmetry of both functions and the bracket $\{\mathcal{H}_{r}, \mathcal{I}_{r}\}=0$ is easily proven by direct computation. Finally, the functional independence of $\mathcal{I}_{r}$ can be easily proven in the harmonic oscillator case $\mathcal{J}=\mathcal{V}_{-2}=0$ with $\la_i=\delta_{ij}$ for a given $j$. Therefore, the functional independence  follows for the generic case with arbitrary $\la_i$ and non-vanishing $\mathcal{J}$ and $\mathcal{V}_{-2}$ functions.

Moreover, the $h_6$-coalgebra symmetry implies that $\mathcal{H}_r$ is not only integrable but superintegrable, since there exists an alternative set of integrals~\cite{alfonsoh6} 
\be
\mathcal{C}_{h_6,(m)}=
{\sum_{ N-m+1\le i<j<k}^{N}} 
\bigl(
\la_i (p_j q_k - p_k q_j ) +
\la_j (p_k q_i - p_i q_k ) +
\la_k (p_i q_j - p_j q_i ) 
\bigr)^2 
\label{c3m}
\ee
where $m=3,\dots,N$. Note that $\mathcal{C}_{h_6,(N)}=\mathcal{C}_{h_{6}}^{(N)}$, so $\mathcal{H}_r$ has a total number of $(2N-4)$ functions in involution (including $\mathcal{I}_r$).

We recall that the ND isotropic oscillator (without any perturbation) has a total number of $(2N-2)$ independent integrals, that make it a maximally superintegrable system. From this perspective, the perturbation given by $\mathcal{F}$ and $\mathcal{G}$ strongly preserves the integrability properties of the oscillator Hamiltonian, since only two integrals of the motion are lost under perturbation as a consequence of imposing the $h_6$ invariance of the system.

\noindent {\bf Remark}. Note that in the case $N=3$ the Hamiltonian (\ref{Hr}) is a particular subclass of the Hamiltonians presented in~\cite{Rauchh6}. This is due to the fact that in our construction we have imposed $\mathcal{V}_{-2}$ to be $h_6$-coalgebra invariant, which is an additional constraint with respect to~\cite{Rauchh6}, where the only condition  imposed on $\mathcal{V}_{-2}$ was to be a homogeneous function of degree $-2$ in the 3D canonical coordinates.

\subsection{A new ND integrable model}
 If we take
\be
\mathcal{V}_{-2}\left(A_{-},  B_{-} \right)=\varepsilon\,\frac{A_{-}^2 - B_{-}}{B_{-}\,(MB_- - A_-^2)}
\ee
and $\la_1=\la_2=\dots=\la_N=1, $
we obtain the Hamiltonian
\be
\mathcal{H}_{r}= \dfrac{1}{2}\sum\limits_{i=1}^{N}\,p_{i}^{2}+\delta_{1}\,\sum\limits_{i=1}^{N}q_{i}^{2}+ 
\varepsilon\,\frac{2\,\sum\limits_{1\leq i<j}^N q_i\,q_j}{\left(\sum\limits_{i=1}^N  q_i^2 \right)\,
\sum\limits_{1\leq i<j}^{N}\left(q_{i}-q_{j}\right)^{2}} +\mathcal{J}\left(\sum\limits_{i=1}^{N}q_{i}^{2}\right)
\label{genCM}
\ee
in which we see that the function $\mathcal{V}_{-2}\left(A_{-},  B_{-} \right)$ gives rise to an homogeneous function of degree $-2$ in the canonical coordinates.
In the $N=2$ case (and with a vanishing radial perturbation $\mathcal{J}$) the quantum mechanical analogue  of (\ref{genCM}) has been studied in~\cite{newCM} as an exactly solvable `generalization' of the 2D Calogero-Moser model. Therefore, presumably (\ref{genCM}) provides a good candidate for a new exactly solvable (classical and quantum) model in arbitrary dimension.

\sect{Another ND integrable perturbation}

Many other possibilities for a perturbed ND oscillator Hamiltonian defined on the $h_6$ Poisson coalgebra can be considered, but only very few special choices admit the existence of an additional integral $\mathcal{I}$. This is the case of the function
\be
\mathcal{H}=\dfrac{1}{2}B_{+}+\delta_{1}\,B_{-}+\mathcal{F}\left({\cal C}_{{\cal G}_{-}}\right)+\mathcal{G}\left(A_{-}\right)
\label{H1gen}
\ee
where the generators of $h_6$ are taken in the ND symplectic realization (\ref{sympn}) and
\be
{\cal C}_{{\cal G}_{-}}=MB_- - A_-^2=
\sum\limits_{1\leq i<j}^{N}\left(\lambda_{j}q_{i}-\lambda_{i}q_{j}\right)^{2}
\ee
is just the Casimir for a (1+1)-Galilei subalgebra of $h_6$ spanned by $\{B_-,A_-,A_+,M\}$. Therefore,  we get the Hamiltonian
\be
\mathcal{H}=\dfrac{1}{2}\sum\limits_{i=1}^{N}\,p_{i}^{2}+\delta_{1}\sum\limits_{i=1}^{N}q_{i}^{2}+ \mathcal{F}\left(\sum\limits_{1\leq i<j}^{N}\left(\lambda_{j}q_{i}-\lambda_{i}q_{j}\right)^{2}\right)
+\mathcal{G}\left(\sum\limits_{i=1}^{N}\,\lambda_{i}q_{i}\right) 
\label{Ham11}
\ee
for which the additional integral $\mathcal{I}$ can also be found in an $h_6$-invariant form through direct computation and reads:
\be
\mathcal{I}=A_{+}^{2}+2\,\delta_{1}\,A_{-}^{2}+2\,\mathcal{G}\left(A_{-}\right)M \notag.
\ee
In terms of the ND symplectic realization, this last integral is explicitly written as
\be
\mathcal{I}=\left(\sum\limits_{i=1}^{N}\,\lambda_{i}p_{i}\right)^{2}\!\!\!\!+2\,\delta_{1}\! \left(\sum\limits_{i=1}^{N}\,\lambda_{i}q_{i}\right)^{2}\!\!
+\, 2\,\left(\sum\limits_{i=1}^{N}\lambda_{i}^{2}\right)\,
\mathcal{G}\!\left(\sum\limits_{i=1}^{N}\lambda_{i}q_{i}\right).
\label{I1}
\ee
Note that $\mathcal{I}$ does not depend on the choice of the function $\mathcal{F}$. 

We stress that the $h_6$-coalgebra symmetry ensures that $\mathcal{I}$ will Poisson-commute with all the $\mathcal{C}_{h_{6}}^{(m)}$ and $\mathcal{C}_{h_6,(m)}$ integrals, and the fact that $\{\mathcal{H},\mathcal{I}\}=0$ is easily proven by direct computation. The functional independence of $\mathcal{I}$ with respect to the remaining integrals is proven by taking $\delta_1=0$ and $\mathcal{G}=0$. In this case  $\mathcal{I}=A_{+}^{2}$ only depends on the momenta, and is indeed functionally independent of $\mathcal{C}_{h_{6}}^{(m)}$ (that does not depend on $\delta_1$, $\mathcal{F}$ and $\mathcal{G}$) and of the Hamiltonian. This functional independence will be mantained under the `deformations' $\delta_1\neq 0$ and $\mathcal{G}\neq 0$.

\subsection {Some examples}

The Hamiltonian (\ref{Ham11}) includes many different systems, and for $N=2,3$ some of them can be found in the literature. For instance, if we consider $\mathcal{F}(x)=\alpha/x$ we get 
\be
\mathcal{H}= \dfrac{1}{2}\sum\limits_{i=1}^{N}\,p_{i}^{2}+\delta_{1}\,\sum\limits_{i=1}^{N}q_{i}^{2}+ 
\frac{\alpha}{\sum\limits_{1\leq i<j}^{N}\left(\lambda_{j}q_{i}-\lambda_{i}q_{j}\right)^{2}} +\mathcal{G}\left(\sum_{i=1}^N \la_i  q_i\right)
\label{CMN}
\ee
which in the case $N=2$ is  a rational Calogero-Moser system plus an arbitrary function $\mathcal{G}$. The integral of the motion in this case is (\ref{I1}) with $N=2$. Note that the integrals (\ref{integrals}) and (\ref{c3m}) only exist from $N=3$ and $\mathcal{I}$ does not depend on the choice of the potential $\mathcal{F}$. 


Nonlinear polynomial oscillators can be obtained from (\ref{Ham11}) through a perturbation 
\be
\phi=\mathcal{F}(x)+\mathcal{G}(y)=\sum\limits_{j=2}^{k}\alpha_{j}\,x^{j} + \sum\limits_{j=3}^{l}\beta_{j-1}\,y^{j},
\label{monomials}
\ee
thus obtaining an integrable superposition of a polynomial deformation of degree $2k$ in the coordinates (the argument $x\equiv {\cal C}_{{\cal G}_{-}}$ is quadratic in $q$)  and another one of degree $l$ (the argument $y\equiv A_{-}$ is linear in $q$). If we analyse separately the homogeneous perturbations with a fixed degree $m$, for odd $m$ only $g(A_{-})$ can contribute, which means that the richest cases will be obtained for $m$ even. 

In particular, the homogeneous quartic case is obtained by taking
$
\phi=\alpha_{2}\,x^{2} + \beta_{3}\,y^{4}.
\label{quartic}
$
In the 2D case, if we take $\lambda_{1}=1,\lambda_{2}=0$ we obtain a separable Hamiltonian.
On the other hand, by taking 
$\lambda_{1}=1, \lambda_{2}=1,\alpha_{2}=\beta_{3}={\epsilon}/{2}$
we recover the well-known $1\!:6\!:\!1$ integrable quartic perturbation (case (4)3' in~\cite{Hietarintacorto} with $E=0$):
\be
\mathcal{H}^{(2)}=\dfrac{1}{2}\left(p_{1}^{2}+p_{2}^{2}\right)+\delta_{1}\left(q_{1}^{2}+q_{2}^{2}\right)+\epsilon \left(q_{1}^{4}+q_{2}^{4}+6q_{1}^{2}q_{2}^{2}\right).
\ee
This system is known to be connected with the Hirota-Satsuma coupled KdV system (see~\cite{BEF}).
In three dimensions, the case $\lambda_{1}=1, \lambda_{2}=0, \lambda_{3}=0$ again provides the separability of the $q_1$ coordinate in the form
\be
\mathcal{H}^{(3)}=\dfrac{1}{2}(p_{1}^{2}+p_{2}^{2}+p_{3}^{2})+\delta_{1}(q_{1}^{2}+q_{2}^{2}+q_{3}^{2})+\alpha_{2}(q_{2}^{2}+q_{3}^{2})^{2}+\beta_{3}q_{1}^{4}
\ee
which is also reflected in the invariant
$
\mathcal{I}=p_{1}^{2}+2(\delta_{1}q_{1}^{2}+\beta_{3}q_{1}^{4})
$. Moreover, the second invariant reads $\mathcal{C}_{h_{6}}^{(3)}=(p_{2}q_{3}-p_{3}q_{2})^2$, which indicates the axial symmetry with respect to the coordinates $(q_2,q_3)$. This is the case (1) of Table 1 in~\cite{Dorizzi}.
On the other hand,  if we take  $\lambda_{1}=\lambda_{2}=\lambda_{3}=1$
we get
\begin{eqnarray}
\mathcal{H}^{(3)}&=&\dfrac{1}{2}\left(p_{1}^{2}+p_{2}^{2}+p_{3}^{2}\right)+\delta_{1}\left(q_{1}^{2}+q_{2}^{2}+q_{3}^{2}\right)\notag \\
&&+2\,\alpha_2\,\left(
(q_{2}-q_{1})^4 + (q_{3}-q_{1})^4+(q_{3}-q_{2})^4
\right) + \beta_3\,(q_{1}+q_{2}+q_{3})^4
\end{eqnarray}
which is just the 3D Chudnowski potential (see~\cite{Rauchh6, Chud}) plus an additional quartic term. 

The generic homogeneous sextic case is obtained by taking 
$
\phi=\alpha_{3}\,x^{3}+\beta_{5}\,y^{6}.
$
In 2D,  separability is trivially obtained when $\lambda_{2}=0$. The case $\lambda_{1}=\lambda_{2}=1$ with $\alpha_{2}=\beta_{5}={b}/{2}$
gives the known coupled sextic oscillator (see~\cite{Hindues}, case (i) of Table 5.3.):
\be
\mathcal{H}^{(2)}=\dfrac{1}{2}\left(p_{1}^{2}+p_{2}^{2}\right)+\delta_{1}\left(q_{1}^{2}+q_{2}^{2}\right)+b\left(q_{1}^{6}+q_{2}^{6}+15\,q_{1}^{2}q_{2}^{2}(q_{1}^{2}+q_{2}^{2})\right).
\ee
In 3D the generic integrable homogeneous sextic perturbation coming from (\ref{Ham11}) gives, under the choice $\lambda_{1}=1,\,\, \lambda_{2}=0,\,\, \lambda_{3}=0$,
a system which is again separable in $q_1$ and whose axial symmetry in $(q_1,q_2)$ is also recovered. As we shall see in the following section, this separability pattern can be explained in general for the Hamiltonian (\ref{Ham11}).


\section{A canonical transformation}

Without any loss of generality, let us assume that $\la_1\neq 0$ and $\la_k\neq 0$ for at least one $k\neq 1$. Let us consider the $N\times N$ orthogonal matrix $R=(a_{i,j})$ with matrix elements 
\begin{eqnarray}
&& a_{1,j}=a_{j,1}=\dfrac{\lambda_{j}}{\lambda}\,\,\,\,\,\qquad j=1,\ldots, N\notag \\
&& a_{i,j}=a_{j,i}=\delta_{i,j}-\dfrac{\lambda_{i}\lambda_{j}}{\lambda(\lambda-\lambda_{1})}\,\,\,\,\qquad (i,j=2,\ldots, N)\notag
\end{eqnarray}
where $\lambda=\sqrt{\sum\limits_{i=1}^{N}\lambda_{i}^{2}}$ and $\delta_{ij}$ is the Kronecker delta. Then $R$ defines a canonical transformation on the (column) vectors $q:=(q_1,q_2,\dots, q_N)$ and $p:=(p_1,p_2,\dots, p_N)$ through
\be
Q=R\cdot q, \qquad  P =R\cdot p
\ee
where
\be
R^{t}=R^{-1}=R, \qquad \mbox{det} R= -1
\ee
and the (column) vectors $Q:=(Q_1,Q_2,\dots, Q_N)$ and $P:=(P_1,P_2,\dots, P_N)$ are the new canonical variables fulfilling $\{Q_{i}, P_{j}\}=\delta_{ij}  $.

In terms of these new variables the $N$-particle symplectic realization of $h_6$ reads
\bea
&& A_+ =\la\, P_1 \qquad\qquad
A_- = \la\,  Q_1\qquad\quad  \kk = \sum_{i=1}^N Q_i P_i -\frac {\la^2}2  \cr 
&& B_+ =\sum_{i=1}^N 
{P_i^2} \qquad\!\quad
 B_- = 
\sum_{i=1}^N  Q_i^2\qquad\,\,
 M = \la^2   ,
\label{sympnQ}
\eea 
and the integrals of the motion coming from the $h_6$ Casimir are transformed into
\be
\mathcal{C}^{(m)}=\la^{2} \,\sum\limits_{2\leq i <j}^{m}(P_{j}Q_{i}-P_{i}Q_{j})^{2} 
\qquad m=3,\ldots,N .
\label{intQP}
\ee

Essentially, the existence of this canonical transformation shows that the symplectic realization  of $h_6$ (\ref{sympn}) with $\la_2=\la_3=\dots=\la_N=0$ is the generic one. In particular, 
under this canonical transformation the Hamiltonian (\ref{Ham11}) reads
\be
\mathcal{H}'=\dfrac{1}{2}\sum\limits_{i=1}^{N}\,P_{i}^{2}+\delta_{1}\sum\limits_{i=1}^{N}Q_{i}^{2}+ \mathcal{F}\left(\la^2\,\sum\limits_{i=2}^{N}\,Q_{i}^{2}\right)
+\mathcal{G}\left(\lambda \,Q_1\right) 
\label{Ham11Q}
\ee
whilst the additional integral (\ref{I1}) is just
\be
\mathcal{I}'=\la^2\left\{ \, P_{1}^{2}+2\,\delta_{1} \, Q_{1}^{2}
+\, 2 \,
\mathcal{G}\!\left(\la\, Q_{1}\right)\right\}.
\label{I1Q}
\ee
Therefore, the $(N-2)$ integrals (\ref{intQP}) just reflect the radial symmetry of the Hamiltonian $\mathcal{H}'$ with respect to the new variables $i=2,\dots,N$, and the remaining one (\ref{I1Q}) expresses the separability of the $i=1$ degree of freedom.

At this point it worth stressing that the canonical transformation $R$ allows us to deduce immediately --and without making use of any $h_6$ invariance--  the existence of the invariants $L_{ij}=(P_{j}Q_{i}-P_{i}Q_{j})$ with $2\leq i<j\leq N$, which are linear in the momenta. Nevertheless, such linear integrals are not generically in involution, and the $h_6$ construction provides automatically the subfamily (\ref{intQP}) of $(N-2)$ quadratic integrals in involution. Anyhow, the linear `angular momentum components' $L_{ij}$ are indeed useful in order to simplify the study of the dynamics of the system and to construct its explicit solutions.

Under the canonical transformation $R$, the Hamiltonian (\ref{Hamr}) is also mapped into
\be
\mathcal{H}_r'=\dfrac{1}{2}\sum\limits_{i=1}^{N}\,P_{i}^{2}+\delta_{1}\,\sum\limits_{i=1}^{N}Q_{i}^{2}
+ \mathcal{V}_{-2}\left(\lambda\,Q_1,  \sum_{i=2}^N  Q_i^2 \right)
+\mathcal{J}\left(\sum\limits_{i=1}^{N}Q_{i}^{2}\right)
\label{HamrQ}
\ee
and its additional integral is written as
\be
\mathcal{I}'_{r}=\sum\limits_{1\leq i <j}^{N}(P_{j}Q_{i}-P_{i}Q_{j})^{2}+2\left(\sum\limits_{i=1}^{N}Q_{i}^{2}\right)\mathcal{V}_{-2}\left(
 {\lambda}\,Q_{1},\sum\limits_{i=1}^{N}Q_{i}^{2}.
\right)
\ee
Note that in general (\ref{HamrQ}) is no longer a separable system, although it does exhibit radial symmetry with respect to the $i=2,3.\dots,N$ degrees of freedom.

\subsection{Rosochatius terms through a new symplectic realization}

After the previous discussion it comes out that the complete integrability of the  Hamiltonian (\ref{Ham11}) is evident when written in the new variables $(Q,P)$.
Nevertheless, this result opens the path to the construction of a non-trivial completely integrable generalization of both (\ref{Ham11Q}) and (\ref{HamrQ}), since if radial symmetry appears  then a Rosochatius-type potential can be added without altering the integrability properties of the Hamiltonian (see~\cite{BH07}, \cite{Annals} and \cite{CRMAngel} for a detailed discussion in terms of the appropriate symplectic realization of the $sl(2)$ Poisson-coalgebra). This enables us to consider a modification of the $N$-particle symplectic realization (\ref{sympnQ}) including a set of $(N-1)$ dimensional Rosochatius terms within the $B_+$ generator as follows:
\bea
&& A_+ =\la\, P_1 \qquad \qquad \qquad
A_- = \la \, Q_1\qquad \qquad    \kk = \sum_{i=1}^N Q_i P_i -\frac {\la^2}2  \cr 
&& B_+ =P_1^2+ \sum_{i=2}^N 
\left( {P_i^2} + \frac{b_i}{Q_i^2}\right)\qquad\!\quad
 B_- = 
\sum_{i=1}^N  Q_i^2\qquad\,\,
 M = \la^2   .
\label{sympnQb}
\eea 
Note that when $b_i>0$ the Rosochatius terms can be properly considered as centrifugal ones. It is immediate to prove that (\ref{sympnQb}) is a symplectic realization for $h_6$, and the Casimir functions under this realization read
\be
\mathcal{C}_b^{(m)}=\la^{2}\left\{ \sum\limits_{i=2}^{m}b_{i}+  \,\sum\limits_{2\leq i <j}^{m}(P_{j}Q_{i}-P_{i}Q_{j})^{2} 
+\sum\limits_{2\leq i < j}^{m}\left(
{b_{j}}\dfrac{Q_{i}^{2}}{Q_{j}^{2}}+{b_{i}}\dfrac{Q_{j}^{2}}{Q_{i}^{2}}\right)\right\}
\qquad m=3,\ldots,N .
\label{intQPb}
\ee

If we now use this symplectic realization for the Hamiltonian (\ref{H1gen}) we get 
\be
\mathcal{H}^b=\dfrac{1}{2}\sum\limits_{i=1}^{N}\,P_{i}^{2}+\delta_{1}\sum\limits_{i=1}^{N}Q_{i}^{2}+ \mathcal{F}\left(\la^2\,\sum\limits_{i=2}^{N}\,Q_{i}^{2}\right)
+\mathcal{G}\left(\lambda \,Q_1\right) + \sum_{i=2}^N 
\frac{b_i}{Q_i^2}
\label{H1QPb}
\ee
which is again completely integrable with constants of the motion given by (\ref{I1Q}) and (\ref{intQPb}).
Note that if the canonical transformation is reversed, the latter Rosochatius terms would be written in a non-trivial way in terms of the original coordinates $(q_1,q_2,\dots, q_N)$. Namely,
\bea
&& Q_{1}^{2}=\dfrac{1}{\lambda^{2}}{\left(\sum\limits_{i=1}^{N}\lambda_{i}q_{i}\right)^{2}}
\\
&& Q_{i}^{2}=\dfrac{1}{\lambda^2(\lambda-\lambda_{1})^2}\left(
(\la-\la_1)(\la_i\,q_1+\la\,q_i)- \la_i\,\sum\limits_{j=2}^{N}\lambda_{j}q_{j}
\right)^{2}\,\,\,\,\,\,i\geq 2.
\eea

In the same way, the symplectic realization (\ref{sympnQb}) can be substituted onto the `abstract' Hamiltonian
$\mathcal{H}_{r}'$ (\ref{Hrabs})
and leads to
\be
\mathcal{H}_{r}^b=
 \dfrac{1}{2}\sum\limits_{i=1}^{N}\,P_{i}^{2}+\delta_{1}\,\sum\limits_{i=1}^{N}Q_{i}^{2}
+ \mathcal{V}_{-2}\left(\la Q_1,  \sum_{i=1}^N  Q_i^2 \right)
+\mathcal{J}\left(\sum\limits_{i=1}^{N}Q_{i}^{2}\right) +  \sum_{i=2}^N 
\frac{b_i}{Q_i^2}.
\ee
This is indeed a new completely integrable Hamiltonian with constants of the motion in involution given by (\ref{intQPb}) and by the symplectic realization (\ref{sympnQb}) of
\be
\mathcal{I}_{r}= B_-B_+-(\kk+\frac 12
M)^2
+2\, B_-\,\mathcal{V}_{-2}\left(A_{-},  B_{-} \right)
\ee
that written in terms of the new canonical coordinates reads:
\bea
&& \!\!\!\!\!\! \!\!\!\!\!\! \!\!\! \!\!\! \!\!\!  \mathcal{I}_{r}^b=\sum\limits_{1\leq i <j}^{N}(P_{j}Q_{i}-P_{i}Q_{j})^{2}+\sum\limits_{i=1}^{N}\sum\limits_{j=2}^{N} b_{j}\dfrac{Q_{i}^{2}}{Q_{j}^{2}}
+2\left(\sum\limits_{i=1}^{N}Q_{i}^{2}\right)\mathcal{V}_{-2}\left(
 {\lambda}\,Q_{1},\sum\limits_{i=1}^{N}Q_{i}^{2}\right)\notag\\
 && \!\!\! \!\!\!  \!\!\! \!\!\! \!\!\! =\dfrac{\mathcal{C}_b^{(N)}}{\la^2}+\sum\limits_{j=2}^{N}
 \left(  (P_{j}Q_{1}-P_{1}Q_{j})^{2}  + b_{j}\dfrac{Q_{1}^{2}}{Q_{j}^{2}}\right)
 +2\left(\sum\limits_{i=1}^{N}Q_{i}^{2}\right)\mathcal{V}_{-2}\left(
 {\lambda}\,Q_{1},\sum\limits_{i=1}^{N}Q_{i}^{2}\right)
\eea
thus showing its functional independence with respect to the integrals $\mathcal{C}_b^{(m)}$ (\ref{intQPb}).

Summarizing, we have presented a constructive approach of two families of completely integrable perturbations-deformations of the ND harmonic oscillator. The fact that all these Hamiltonians  can be written as functions of the $h_6$-Poisson coalgebra under suitable $N$-particle symplectic realizations  makes it possible to demonstrate the (super)integrability properties of both classes of systems. 
Among all the Hamiltonians here presented, the model (\ref{genCM}) seems to deserve some attention. In particular, its explicit solvability could be addressed through the `cluster variables' technique~\cite{cluster} in order to select the appropriate dynamical variables by making use of the underlying coalgebra structure.  Another interesting possibility for further work would be the construction of the analogues on ND spaces with constant curvature of the Hamiltonians here presented. This could be feasible by mimicking the $sl(2,\mathbb R)$ approach described in~\cite{BH07,PhysD,Annals} but now making use of the corresponding $h_6$-coalgebra formalism.

\section*{Acknowledgements}

\small

This work was partially supported by the Spanish MICINN  under grant  MTM2007-67389 (with EU-FEDER support), by Junta de Castilla y
Le\'on  (Project GR224) and by INFN-CICyT. A. Blasco acknowledges INFN support during his stay in Roma Tre University, where part of this work was completed. The authors acknowledge O. Ragnisco and F.J. Herranz for helpful suggestions and one of the referees for suggesting the existence of the canonical transformation described in Section 4.



\begin{thebibliography}{40}



\bibitem{Perelomov} A.M. Perelomov, {\it Integrable Systems
of Classical Mechanics and Lie algebras}, Berlin: Birkh\"auser (1990).

\bibitem{Fradkin}
 D.M. Fradkin, 
{\em Am. J. Phys.}
{\bf 33}  207  (1965).



\bibitem{Gutzwiller} M.C. Gutzwiller, {\it Chaos in Classical and Quantum Mechanics}, Interdisciplinary Applied Mathematics, Vol. 
1, Springer, New York (1990).

\bibitem{Hietarinta}
J. Hietarinta, 
{\it Phys. Rep.} {\bf 147}  87  (1987).

\bibitem{RDGprl} A. Ramani, B. Dorizzi, B. Grammaticos,   
{\it Phys. Rev. Lett.} {\bf 49} 1539 (1982).

\bibitem{GDP}
 B. Grammaticos, B. Dorizzi, R. Padjen, 
{\em Phys. Lett. A}
{\bf 89}  111  (1982).

\bibitem{GDR} B. Grammaticos, B. Dorizzi, A. Ramani, 
{\it J. Math. Phys.} {\bf 24} 2289 (1983).

\bibitem{Hietarintacorto} J. Hietarinta, 
{\em Phys. Lett. A}
{\bf 96}  273  (1983).

\bibitem{Rauchh6}
S.  Wojciechowski,
{\it Phys. Lett. A} {\bf 96} 389 (1983).

\bibitem{woja}
 M. Wojciechowska,
{\it J. Phys. A: Math. Gen.} {\bf 17} 1993 (1984).


\bibitem{Dorizzi} B. Dorizzi, B. Grammaticos, J. Hietarinta, A. Ramani, F. Schwarz,
{\it Phys. Lett. A} {\bf 116} 432 (1986).

\bibitem{Palacian}
 J. Palaci\'an and P. Yanguas, 
{\em Chaos, Solitons and Fractals.}
{\bf 11}  879  (2000).

\bibitem{MRW}
K. Marciniak, S. Rauch-Wojciechowski,
{\it Inverse Problems} {\bf 17} 191 (2001).
   
\bibitem{MRWreports}
K. Marciniak, S. Rauch-Wojciechowski,
{\it Rep. Math. Phys.} {\bf 48} 139 (2001).


\bibitem{RamaniNuevos} A. Ramani, J. Hietarinta, B. Dorizzi, B. Grammaticos, 
{\it Phys. Lett. A} {\bf 109} 55 (1985).

\bibitem{2toN} B. Grammaticos, B. Dorizzi, A. Ramani, J. Hietarinta,   
{\it Phys. Lett. A} {\bf 109} 81 (1985).


\bibitem{Hindues} M. Lakshmanan and R. Sahadevan, {\it Phys. Rep.} {\bf 224} 1 (1993).

\bibitem{BruschiC}
M. Bruschi and F. Calogero,
{\it Phys. Lett. A} {\bf 273} 173  (2000).
   


\bibitem{Garnier}
 R. Garnier,
{\it Rend. Circ. Math. Palermo} {\bf 43} 155 (1919).

 \bibitem{fris}
   J.  Fris, V. Mandrosov, Ya A. Smorodinsky, M. Uhlir and P. Winternitz,
{\it   Phys. Lett.} {\bf 16}  354 (1965).

 \bibitem{evans2}
  N.W.   Evans,
{\it   Phys. Lett. A} {\bf 147}  483 (1990).

\bibitem{Ca71}
F. Calogero,
{\it J. Math. Phys.} {\bf 12} 419 (1971).

\bibitem{Mo75}
J. Moser,
{\it Adv. Math.} {\bf 16} 197 (1975).






\bibitem{FWM}
A. Fordy, S. Wojciechowski and I. Marshall,
{\it Phys. Lett. A} {\bf 113} 395 (1986).
   

\bibitem{BH07}
A. Ballesteros, F.J. Herranz,
{\em J. Phys. A:  Math. Theor.}
{\bf 40}  F51  (2007).



 \bibitem{PhysD}
A. Ballesteros, A. Enciso, F.J. Herranz, O. Ragnisco,
 {\it Physica D}   {\bf 237} 505 (2008).

\bibitem{Annals}
A. Ballesteros, A. Enciso, F.J. Herranz, O. Ragnisco,
{\it Ann. Phys.} {\bf 324} 1219 (2009).

   



\bibitem{BR} A. Ballesteros, O. Ragnisco, {\it J.
Phys. A: Math. Gen.} {\bf 31} 3791 (1998).

\bibitem{BHnonlin} A. Ballesteros, F.J. Herranz,  
{\it J. Nonlin. Math. Phys.} {\bf 8} Suppl. 18 (2001).

\bibitem{alfonsoh6}
  A. Ballesteros, A. Blasco, F.J. Herranz,
  {\it J. Phys. A: Math. Theor.} {\bf 42}, 265205 (2009). 

 \bibitem{CRMAngel}
A. Ballesteros, A. Blasco, F.J. Herranz, F. Musso, O. Ragnisco,
  {\it J. Phys.: Conf. Ser.} {\bf 175}, 012004 (2009). 


\bibitem{Rosochatius1}
 E. Rosochatius, 
 {\it \"Uber die Bewegung eines Punktes}, Inaugural 
Dissertation, Universit\"at Gottingen (Gebr. Unger, Berlin)
(1877).

\bibitem{Rosochatius2}   T. Ratiu, {\it The Lie algebraic interpretation of the complete 
integrability of the Rosochatius system}, in: Mathematical 
Methods in Hydrodynamics and Integrability in Dynamical 
Systems, AIP Conference Proc., Vol. 88 (AIP, New York) p. 109 
(1982). 


\bibitem{Gilmore}
 W. M. Zhang,  D. H. Feng, R. Gilmore, 
 {\it Rev. Mod. Phys.} {\bf 62}   867  (1990).

\bibitem{Patera}   G. Burdet, J. Patera, M. Perrin and
P. Winternitz, {\it J. Math. Phys.} {\bf  19}  1758 (1978).



\bibitem{newCM}
A. Diaf, A. T. Kerris, M. Lassaut, R.J. Lombard,
{\it J. Phys. A: Math. Gen.} {\bf 39} 7305 (2006).

\bibitem{BEF}
S. Baker, V.Z. Enolskii and A.P. Fordy,
{\it Phys. Lett. A} {\bf 201} 167 (1995).


\bibitem{Chud}G. V. Chudnovsky, D. V. Chudnovsky,
{\it Lett. Nuovo Cimento} {\bf 19} 291 (1977).





\bibitem{cluster}
A. Ballesteros, O. Ragnisco,
{\em J. Phys. A:  Math. Gen.}
{\bf 36}  10505  (2003).








\end{thebibliography}
\end{document}